\documentclass[a4paper,11pt]{article}
\pdfoutput=1 

\usepackage{jheppub} 
\usepackage{hyperref}
\usepackage[T1]{fontenc} 
\usepackage[normalem]{ulem}

\usepackage{graphicx}
\usepackage{dcolumn}
\usepackage{bm}

\usepackage{color,xcolor,url,ulem,verbatim,appendix}

\newcommand{\beq}{\begin{equation}}
\newcommand{\eeq}{\end{equation}}
\newcommand{\bea}{\begin{eqnarray}}
\newcommand{\eea}{\end{eqnarray}}

\newcommand{\ep}{\epsilon}

\newcommand{\mpl}{M_{\text{Pl}}}

\pdfoutput=1 

\title{\begin{center}
    \LARGE{\textbf{Cosmological Phase Transition of Spontaneous Confinement}}
\end{center}
}
\author[a]{Kaustubh Agashe}
\emailAdd{kagashe@umd.edu}
\author[a,b]{Peizhi Du}
\emailAdd{peizhi.du@stonybrook.edu}
\author[a]{Majid Ekhterachian}
\emailAdd{ekhtera@umd.edu}
\author[a]{Soubhik Kumar}
\emailAdd{soubhik@terpmail.umd.edu}
\author[a]{Raman Sundrum}
\emailAdd{raman@umd.edu}
\affiliation[a]{Maryland Center for Fundamental Physics, Department of Physics,\\ University of Maryland, College Park, MD 20742 USA}
\affiliation[b]{C.N. Yang Institute for Theoretical Physics, Stony Brook University, Stony Brook, NY, 11794, USA}

\abstract{
The dynamics of a cosmological (de)confinement phase transition is studied in nearly conformally invariant field theories, where confinement is predominantly spontaneously generated and associated with a light ``dilaton'' field. We show how the leading contribution to the transition rate can be computed within the dilaton effective theory. In the context of Composite Higgs theories, we demonstrate that a simple scenario involving two renormalization-group fixed points can make the transition proceed much more rapidly than in the minimal scenario, thereby avoiding excessive dilution of matter abundances generated before the transition. The implications for gravitational wave phenomenology are discussed. In general, we find that more (less) rapid phase transitions are associated with weaker (stronger) gravitational wave signals. The various possible features of the strongly coupled composite Higgs phase transition discussed here can be concretely modeled at weak coupling within the AdS/CFT dual Randall-Sundrum extra-dimensional description, which offers important insights into the nature of the transition and its theoretical control. These aspects will be presented in a companion paper.
}

\begin{document}
\hspace{23em} UMD-PP-019-05, YITP-SB-19-32
\maketitle
\flushbottom
\section{Introduction} First order phase transitions (PT) can play an important role in cosmological evolution through  dramatic rearrangements of particle physics degrees of freedom (d.o.f).
As out-of-equilibrium processes, such PTs can create new matter asymmetries, or drastically alter pre-existing ones. They also can provide a spectacular source for stochastic gravitational waves (GW) \cite{Kosowsky:1991ua,Kosowsky:1992vn,Kosowsky:1992rz,Kamionkowski:1993fg} (see reference \cite{Caprini:2015zlo} for a review). While the  Standard Model (SM) Higgs boson does not give rise to a first order electroweak (EW) PT (see reference \cite{Quiros:1999jp} for a review), this PT can be first order in many beyond-SM (BSM) extensions. Further, BSM extensions may give rise to other PTs, roughly connected to the EW scale by the naturalness principle. There may then be one or more PTs in the $\sim$ TeV - $100$ TeV range.  If so, we should be able to probe such BSM physics by complementary means,  its microphysics at particle collider experiments, and the associated PT in  GW detectors \cite{Audley:2017drz,Harry:2006fi,Graham:2017pmn,Kawamura:2011zz,Gong:2014mca}. 

Theories in which the Higgs boson is a tightly confined composite made of more fundamental constituents (see references \cite{Bellazzini:2014yua,Panico:2015jxa} for a review) are particularly promising in this regard, naturally generating realistically large particle physics hierarchies. Beyond the EW PT itself, they undergo a fascinating and rich PT between the confined and deconfined phases,  albeit a non-perturbative and theoretically challenging one. While a (de)confinement PT is not necessarily first order, as illustrated by QCD for realistic parameters \cite{Petreczky:2012rq}, it can readily be. Composite Higgs flavor-physics typically requires strong coupling over a large hierarchy of scales, such as occurs in the domain of an approximate fixed point (FP) of the renormalization group (RG), and plausibly a large-$N(\text{color})$ structure (see reference \cite{Coleman:1985rnk} for a review of large $N$).  Greater theoretical control of the strong dynamics is then possible if the large-$N$ approximate FP conformal field theory (CFT) has a useful Anti-de Sitter (AdS)/CFT dual description \cite{Maldacena:1997re,Witten:1998qj,Aharony:1999ti}. Indeed, most of the realistic model building has been done in such a dual higher-dimensional Randall-Sundrum (RS) warped spacetime \cite{Randall:1999ee,Randall:1999vf,ArkaniHamed:2000ds,Rattazzi:2000hs} (see \cite{Davoudiasl:2009cd,Gherghetta:2010cj} for reviews).

Another controlled regime, already visible in four dimensional (4D) spacetime without AdS/CFT, occurs if the breaking of approximate conformal invariance by confinement is primarily {\it spontaneous}, resulting in a light pseudo Nambu-Goldstone boson (PNGB) ``dilaton'' field $\phi$ \cite{PhysRev.184.1760}. Here, the vacuum expectation value (VEV) $\langle\phi\rangle$ gives the confinement scale which typifies the masses of generic composites. This structure was first seen in composite Higgs theory in the dual RS formulation in terms of the ``radion'' \cite{Randall:1999ee,Randall:1999vf,Goldberger:1999uk}.
One goal of this paper is to re-analyse the PT using the 4D dilaton effective field theory (EFT) \cite{Rattazzi:2000hs,Goldberger:2008zz,Chacko:2012sy,Bellazzini:2012vz,Chacko:2013dra,Coradeschi:2013gda,Chacko:2014pqa}
and reasonable physical expectations, as far as possible. In particular, we study the conditions under which the dilaton dynamics dominates the bubble nucleation rate, which competes with the cosmological expansion rate. Ultimately, a fuller description and justification of these expectations involves modeling the deconfined phase, outside the dilaton regime, a task we will re-examine in a forthcoming paper from the RS perspective \cite{5dpaper}. This dual description requires large $N$ and yields a more tractable semi-classical, but higher-dimensional description of non-perturbative 4D deconfinement in terms of the AdS-Schwarzschild horizon. The confinement PT then corresponds to bubbles of the RS ``IR brane'' nucleating and expanding from this horizon \cite{Creminelli:2001th}.  Our 5D analysis \cite{5dpaper} will further justify and sharpen the dilaton dominance approximation and account for subleading corrections. Therefore, here, we will track the consistency of our dilaton dominance results with large $N$.

Reference \cite{Creminelli:2001th} already argued for dilaton dominance in the RS context, but not completely within higher-dimensional EFT control, and they showed that the PT cannot be prompt in the minimal RS model. References \cite{Randall:2006py,Kaplan:2006yi,Nardini:2007me} showed that the PT could nevertheless complete after a period of supercooling, assuming dilaton dominance (see also \cite{Konstandin:2011dr,vonHarling:2017yew,Bruggisser:2018mus,Bruggisser:2018mrt,Baratella:2018pxi} for further studies of supercooling). Our results will reinforce the earlier work more systematically. Furthermore, we will also show that having {\it separate} approximate RG FP regimes controlling large hierarchies and the PT dynamics can easily result in a more prompt PT than the minimal model, with important consequences for cosmological (dark) matter abundances, GW and collider phenomenology. References \cite{Hassanain:2007js,Konstandin:2010cd,Dillon:2017ctw,Bunk:2017fic,Megias:2018sxv} explored other non-minimal modeling to make the PT complete more promptly.  

This paper is organized as follows. In Section \ref{sec.eq}, we give the equilibrium description of the confined and deconfined phases, and then in Section \ref{sec.thin}, we calculate the rate of the phase transition between the two phases in the thin-wall regime. We notice that in the minimal composite Higgs models where the Planck-Weak hierarchy is correctly accounted for, the PT does not complete in the thin-wall regime if we demand a theoretically controlled analysis. In such cases, the universe supercools for a very long time and dilutes any pre-existing particle abundances. Therefore in Section \ref{sec.twofp}, we construct a simple modification of the minimal scenario involving two non-trivial fixed points. In this modified scenario, although the PT has a better chance of completing within the thin-wall regime, it will still often complete only after some supercooling. However, we show in Section \ref{sec.super}, that the extent of supercooling need only be mild, and thus any pre-existing abundances do not get significantly diluted. After discussing the associated gravitational wave signatures in Section \ref{sec.gw}, we conclude in Section \ref{sec.con}. 

\section{Equilibrium description of the two phases}\label{sec.eq}
We model the deconfined phase as an approximate CFT, coupled to gravity, with $\mathcal{O}(N^2)$ d.o.f. At a temperature $T$, its free energy (density) $F$ can be written as \cite{Creminelli:2001th},
\begin{equation}\label{eq:F_decon}
F_{\rm deconfined}= V_0-C N^2 T^4,
\end{equation}
where $V_0$ is a vacuum energy in the deconfined phase and $C$ is some strong-coupling model-dependent $\mathcal{O}(1)$ constant. 
At low enough $T$ the theory can spontaneously confine giving rise to massive composite states. One of the light composites will be the PNGB dilaton, as noted above.
In addition, there may be an $\mathcal{O}(1)$ number of other light composites, in particular the composite Higgs boson, which are weakly coupled to the dilaton by $1/N$. However, it is the dilaton that will play the central role in determining the bubble nucleation rate, as discussed below.
We will therefore neglect the other light composites. Further there may be other light elementary particles. They are very weakly coupled to the dilaton, are present in both phases, and are essentially spectators to the PT.

Below the spontaneous confinement scale, we work in the  dilaton EFT.   We model the small departure from conformal invariance by $\Delta\mathcal{L} = g \mathcal{O}$,  where $\mathcal{O}$ is a nearly-marginal composite operator and where the coupling $g$ runs from the UV,  but stops at the confinement scale, locally given by $\phi(x)$. This is the only way in which conformal invariance is broken within the compositeness dynamics, leading to an effective Lagrangian :
\begin{equation}\label{dilatonpot}
\mathcal{L}_{\text{eff}}=\frac{N^2}{16\pi^2}\left((\partial\phi)^2-\lambda \left(  g( \phi) \right) \phi^4\right)-V_0,   
\end{equation}
where the explicit breaking is characterized by the ``running'' quartic coupling $\lambda \left(  g(\phi) \right)$. We see that if $g$ did not run, the dilaton coupling would be exactly conformally invariant $\phi^4$.
The vacuum energies of the two phases are equated by matching at the common limit of the two phases, $T=0$, $\phi=0$. This vacuum energy also breaks conformal invariance but is only of gravitational relevance.
In this standard large-$N$ ``glueball'' normalization (reviewed in \cite{Coleman:1985rnk,Manohar:1998xv}), the self-coupling is expected to be $\lambda\sim 1$. However,  it is certainly possible that $\lambda$ is somewhat smaller, in which case theoretical control can be gained by expanding in $\lambda$, as we will see below.

For a small deformation $g$, we can expand $\lambda$ to first order,
\begin{equation}\label{grunning}
\lambda(g)=\lambda_0+\lambda'_0 g,
\end{equation}
where $\lambda_0\equiv\lambda(g=0)$ and $\lambda'_0 \equiv \frac{d \lambda}{dg}|_{g=0}$.
For  $\beta(g)\equiv\frac{dg}{d\ln \mu} \approx \epsilon g$, the scaling dimension of $\mathcal{O}$ is determined to be $4+\epsilon$, and $g(\phi) \approx g_{\text{UV}} \left(\frac{\phi}{\Lambda_{\text{UV}}}\right)^{\epsilon}$,
where $g_{\text{UV}}$ is the deformation at UV cut-off scale $\Lambda_{\text{UV}}$. Plugging this and eq. \eqref{grunning} into eq. \eqref{dilatonpot}, gives us the explicit form for the 
leading dilaton potential from which we derive the confinement scale,
\begin{equation}\label{hierarchy}
\langle \phi \rangle=\Lambda_{\text{UV}} \left( -\frac{1}{1+\epsilon/4}\frac{\lambda_0}{\lambda'_0 g_{\text{UV}}}  \right)^{\frac{1}{\epsilon}}.
\end{equation}
We note that an exponentially large hierarchy between $\langle \phi \rangle$ and $\Lambda_{\text{UV}}$ can be obtained if $\epsilon$ is small, given just a mild hierarchy between $\lambda_0$ and $\lambda^\prime_0g_{\text{UV}}$ \cite{Rattazzi:2000hs}. This is dual to the minimal 5D Goldberger-Wise mechanism \cite{Goldberger:1999uk}. It is convenient to express the potential in terms of $\langle\phi\rangle$,
\begin{equation}\label{radionpot1}
V_{\text{eff}} =\frac{N^2}{16\pi^2} \lambda_0\phi^4 \left(1-\frac{1}{1+\epsilon/4} \left( \frac{\phi}{\langle\phi\rangle}\right)^{\epsilon}   \right) +V_0.
\end{equation}
We choose $V_0$ to ensure the (almost) vanishing cosmological constant (CC) today, i.e. we impose $V_{\text{eff}}(\langle \phi \rangle)=0$. Note that, vacuum stability implies $\epsilon\lambda_0 <0$.

Assuming a low critical temperature for the PT, $T_c \ll \langle\phi\rangle$, we can solve for it by equating the free energies of the two phases:
\begin{align}\label{Tc}
&F_{\rm deconfined}(T_c)=F_{\text{confined}}(T_c)\underset{{T_c\ll \langle \phi \rangle}}{\approx} V_{\text{eff}}(\langle\phi\rangle) \nonumber\\&\Rightarrow  
\frac{T_c}{\langle\phi\rangle}= \left(\frac{-\ep \lambda_0}{16\pi^2C(4+\ep)}\right)^{1/4}+\mathcal{O}\left(\frac{1}{N^2}\right).
\end{align}
We see that $T_c$ is self-consistently small for small $\ep$ and/or small $\lambda_0$. Therefore the confining phase is within dilaton EFT control. Since the coupling $g(\phi)$ blows up in the IR for $\epsilon<0$, making the bounce calculation unreliable, we will consider $\epsilon>0,\lambda_0<0$ in the minimal set-up. With this choice, approximate conformal invariance only improves in the IR, so that the
deconfined phase is expected to exist at arbitrarily small $T$, including at $T_c$. This expectation is borne out in the dual RS analysis \cite{Creminelli:2001th,5dpaper}. The simultaneously allowed phases at $T_c$ indicate a first-order PT. It follows from eq. \eqref{Tc} that $V_0=CN^2T_c^4$ for vanishing CC today.

A cosmological PT completes for sufficiently large bubble nucleation rate per unit volume, $\Gamma\geq H^4$, where $H$ is the Hubble scale. For $T<T_c$, $H$ 
asymptotes to a constant, driven
by vacuum energy, $H^2\approx \frac{8\pi }{3} G_N V_0 \sim\frac{C N^2 T_c^4}{3M_{\text{Pl}}^2}$. Here, $G_N$ and $M_{\text{Pl}}$ are respectively Newton's constant and the reduced Planck scale, $M_{\text{Pl}}=2.4\times 10^{18}$ GeV. Semi-classically the finite temperature bubble nucleation rate $\Gamma$, is computed in terms of the Euclidean bounce action $S_{\text{b}}$ with time periodicity $1/T$ as,
\begin{equation}\label{Gamma}
\Gamma \sim T^4e^{-S_{\text{b}}}\underset{\text{completion}}{\geq} H^4.
\end{equation}
Thus for the PT to complete, $S_{\text{b}} < 4\ln\left(\frac{\mpl}{T_c}\right)\sim 140$ for $T_c\sim$ TeV. For small $\lambda$, as we would expect, and will show in the Appendix, the dominant finite-temperature bounce solutions are $O(3)$ symmetric (and Euclidean time independent). 

\section{Phase transition in the thin-wall regime} \label{sec.thin}
Let us first compute $\Gamma$ in the thin-wall approximation, for prompt PT, $T\approx T_c$. In this approximation quite generally \cite{Coleman:1977py,Linde:1981zj}
\begin{equation}
S_\text{b}=\frac{S_3}{T} = \frac{16 \pi}{3}\frac{S_1^3}{(\Delta F)^2 T},
\end{equation}
where  $\Delta F$ is the free energy difference between the two phases and $S_1$ is the surface tension of the bubble wall. The bubble has to interpolate between the de-confined and the confined phases, see Fig. \ref{fig:dilatonpot}. This interpolation consists of two regions, (i) the lowering of the dynamical confinement scale from $\langle\phi\rangle$ down to $\sim T_c\ll \langle\phi\rangle$, followed by (ii) the rearrangement of all d.o.f from confined into deconfined at $\lesssim T_c$ scales. The first region is described purely within the dilaton EFT. To see this note that the dilaton bounce solutions have $|\nabla\phi|\sim \sqrt{V_\text{eff}}$ which implies $|\nabla\phi|/\phi^2\sim\sqrt{|\lambda(g)|}\ll 1$ for small $\lambda_0$. Thus for $\phi>T_c$, gradients and $T$ are smaller than the local mass gap $\phi$, and do not excite the heavier composite d.o.f. In this dilaton dominance approximation we find
\begin{align}\label{S1thin}
    S_1^{\text{(i)}}\approx \frac{N}{2 \pi}\int _{\sim T_c}^{\langle\phi\rangle}d\phi \sqrt{ V_{\text{eff}}}\approx 0.6 \Big(\frac{C^3}{\epsilon |\lambda_0|} \Big)^{1/4} N^2 T_c^3.
\end{align}
We see this is enhanced by small $\epsilon$ and $\lambda_0$ in $T_c$ units because $\phi$ is getting large in these units over the bounce trajectory as seen from eq. \eqref{Tc}. We are therefore insensitive to the lower limit of integration which we can approximate as vanishing. In region (ii), $\phi/T_c\sim \mathcal{O}(1)$ so that we do not expect enhancement by small $\epsilon$ or $\lambda_0$. Therefore we have dilaton dominance, $S_1\approx S_1^{\text{(i)}} $, 
\begin{equation}\label{S3thinwall}
\frac{S_3}{T} \approx 3.6 
\left(\frac{1}{|\lambda_0|\epsilon}\right)^{\frac{3}{4}}C^{\frac{1}{4}} N^2\frac{T_c/T}{\left(1-(T/T_c)^4\right)^2}.
\end{equation}

Let us apply the above result to the case of a PT at very roughly TeV scale in the minimal scenario in which $\epsilon$ accounts for the Planck-TeV hierarchy,  $\epsilon\approx 1/25$. But we see from eq. \eqref{S3thinwall} that a prompt PT cannot occur within theoretical control, even for $|\lambda_0|=1/2,\frac{T^4_c-T^4}{T_c^4}=1/2$ and $N>1$! To allow the PT to happen for larger values of $N$, we need larger values of $\epsilon$ while still somehow generating a large hierarchy. We now describe a simple scenario which achieves that.

\section{A two-FP RG evolution}\label{sec.twofp}
Earlier, to obtain eq. \eqref{radionpot1} we approximated $\beta(g)\approx \epsilon g$ for near-FP behaviour. 
 However, it is possible that the running flows to this vicinity from a different UV FP at $g_*$.
We then have two important critical exponents:
\begin{equation}
\beta(g)=
\begin{cases}
\epsilon' (g_*-g)   &\text{for } g \text{ near } g_*\\
\epsilon g  & \text{for  $g$ small}.
\end{cases}
\end{equation}
The transition between the two regimes happens around some intermediate coupling, $g\sim g_{\text{int}}$ at a scale $\Lambda_{\text{int}}\sim \Lambda_{\text{UV}} \left(\frac{g_*-g_{\text{UV}}}{g_*-g_{\text{int}}}\right)^{1/\epsilon'}$. The confinement scale is now generated from $\Lambda_{\text{int}}$ analogously to eq. \eqref{hierarchy} but with replacements $\Lambda_{\text{UV}}\rightarrow \Lambda_{\text{int}}$ and $g_0\rightarrow g_{\text{int}}$,
\begin{equation}
\langle\phi\rangle \sim \left(\frac{g_*-g_{\text{UV}}}{g_*-g_{\text{int}}}\right)^{1/\epsilon'} \left(-\frac{\lambda_0}{(1+\epsilon/4)\lambda'_0g_{\text{int}}}\right)^{1/\epsilon}\Lambda_{\text{UV}}.
\end{equation}
We see that we can now have a larger $\epsilon$ controlling the PT dynamics while still having a large Planck-TeV hierarchy given by small $\epsilon^\prime$ (for a related idea see \cite{Baratella:2018pxi}). Eq. \eqref{radionpot1} implies that the dilaton $\text{mass}^2\propto\epsilon$, and hence 
a larger $\epsilon$ implies a heavier dilaton relative to the confinement scale $\langle \phi \rangle$, relevant for collider searches. The above two-FP structure of RG running can be simply modeled with a suitable 5D scalar potential in the dual RS formulation \cite{5dpaper}. By contrast, the standard Goldberger-Wise 5D scalar \cite{Goldberger:1999uk} with only a mass term in the RS ``bulk'' is dual to the minimal scenario discussed above.

For a benchmark set of parameters $\epsilon = 0.5,|\lambda_0|=0.5, C=1,\frac{T^4_c-T^4}{T_c^4}=1/2$, the bounce action can be obtained using eq. \eqref{S3thinwall}, with eq. \eqref{Gamma} showing that the PT can complete promptly for $N\approx 2$. This is marginally in theoretical control. If we are outside the regime/parameters for prompt PT, the universe remains and cools in the deconfined phase, and inflates due to the constant term in eq. \eqref{eq:F_decon}. Ultimately, the PT may complete in a supercooled regime, $T \ll T_c$. We now turn to this analysis.

\section{Phase transition in the supercooled regime} \label{sec.super}
\begin{figure}
    \centering
    \includegraphics[scale=0.6]{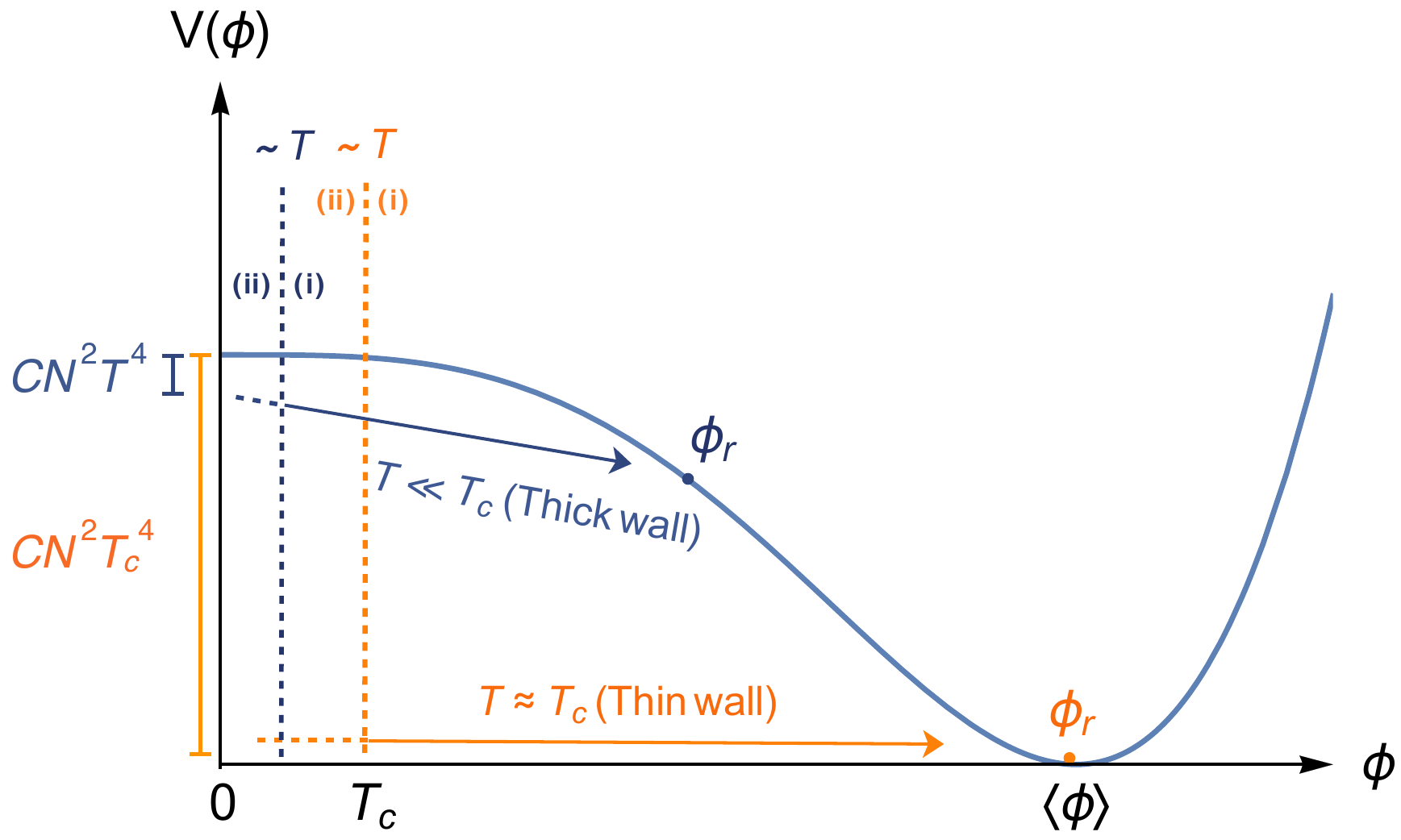}
    \caption{The scalar field dynamics of the PT in the prompt (orange) and supercooled (blue) regimes. $\phi_r$ denotes the release point, characterizing the value of the dilaton field at the center of the bubbles at the moment of their nucleation. The free energy in the deconfined phase is indicated along the vertical axis. The parts of the bounce trajectory to the right/left of the vertical dotted lines correspond to regimes (i)/(ii) in the text giving dominant/sub-dominant contributions to $S_{\text{b}}$.}
    \label{fig:dilatonpot}
\end{figure}
For $T\ll T_c$, by eqs. \eqref{eq:F_decon} and \eqref{radionpot1} 
the release point in $\phi$ drops \cite{Konstandin:2010cd}, see Fig. \ref{fig:dilatonpot} . Therefore the bounce only probes
the dilaton potential for small $\phi$, $V_{\text{eff}}\approx\frac{N^2}{16\pi^2}\lambda_0\phi^4+V_0$. In this regime we can use a scaling argument for the $O(3)$ symmetric bounce action,
\begin{align}\label{scaledactionfull}
\frac{S_{3}^{\text{(i)}}}{T} =& \frac{N^2}{4\pi T}\int dr r^2 \left( \left(\frac{d\phi}{dr}\right)^2+\lambda_0\phi^4 +16 \pi^2 CT^4\right)\\ \label{scaledaction}
=& \frac{N^2}{4\pi |\lambda_0|^{\frac{3}{4}}}\int dx x^2 \left(\left(\frac{d\tilde{\phi}}{dx}\right)^2-\tilde{\phi}^4+ 16 \pi^2 C\right),
\end{align}
where $\tilde{\phi}=|\lambda_0|^{\frac{1}{4}}\phi/T$ and $x = |\lambda_0|^{\frac{1}{4}} rT$.  Thus we see that the $S_{3}^{\text{(i)}}$ is not enhanced by $\epsilon$ compared to thin wall eq. \eqref{S3thinwall}, allowing a larger nucleation rate at low $T$.  The dilaton profile is then given by extremizing this action  subject to two boundary conditions (BC). One is given by $\frac{d\phi}{dr}=\frac{d\tilde{\phi}}{dx}=0$ at $r=0$. For the other BC, we first note that part (ii) of the bounce connects to part (i) for $\phi\sim T\ll T_c \ll \langle\phi\rangle$ which we approximate as $\phi\approx 0$ i.e. $\tilde{\phi}\approx 0$. Due to the fact that part (ii) of the bounce is insensitive to small $|\lambda_0|\ll 1$, we will have  a $\lambda_0$-independent kinetic/gradient energy $(\frac{d\phi}{dr})^2|_{\phi=0} =T^4(\frac{d\tilde{\phi}}{dx})^2|_{\tilde{\phi}=0}$ where $(\frac{d\tilde{\phi}}{dx})^2|_{\tilde{\phi}=0}$  is some $\mathcal{O}(1)$ number which we will fix below. These BCs imply that $\tilde{\phi}(x)$ is independent of $\lambda_0$ and therefore the radius of the bubble where $\phi\approx 0$ is $\propto \frac{1}{|\lambda_0|^{1/4}}$. Beyond this radius, the $\lambda_0$-independent physics of part (ii) forms a ``thin-wall'' $\sim (\lambda_0)^0$ around the larger part (i) of the profile. Thus, $S_{3}^{\text{(ii)}}$ is proportional to the area of the bubble $\propto\frac{1}{|\lambda_0|^{1/2}}$. 
The gradient energy at the matching point $\phi\approx 0$ is then given by the thin-wall approximation $(d\phi/dr)^2\approx \frac{16 \pi^2}{N^2} \Delta F= 16 \pi^2 C T^4$.
To summarize, $S_{3}^{\text{(i)}}\propto |\lambda_0|^{-3/4}$ while $S_{3}^{\text{(ii)}}\propto|\lambda_0|^{-1/2}$,  demonstrating dilaton dominance for $|\lambda_0|\ll 1$. In reference \cite{5dpaper} we will quantify and include the next-to-leading contribution due to region (ii).

\begin{figure}[h]
\centering
\includegraphics[width=0.6\linewidth]{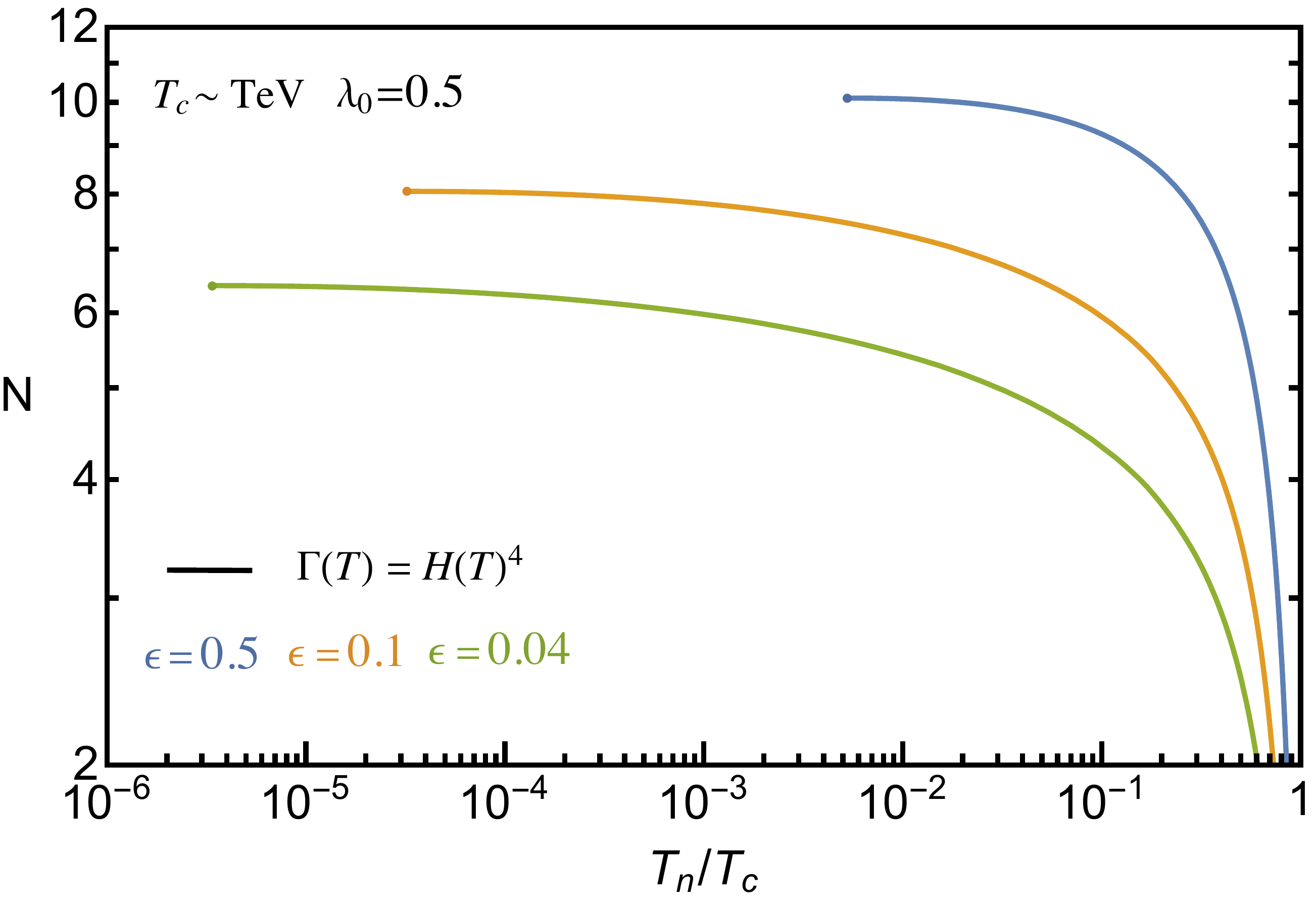}
\caption{Lines depicting the nucleation temperature as a function of $N$ for different choices of $\epsilon$, obtained by saturating eq. \eqref{eq:action_max} . The interpretation can be found in the text. As explained in the text, the inverse duration of the PT, $\beta_{\text{GW}}$, is proportional to the slope of these curves. The end points of the lines denote the maximum allowed value of $N$ for which the PT completes for given $\epsilon,\lambda_0$.}
\label{fig:nt}
\end{figure}

Having demonstrated dilaton dominance for extreme $T$, 
we expect it to hold for all $T$, in particular, intermediate temperatures. We then evaluate the bounce action numerically with the BC above. The results are shown in Fig. \ref{fig:nt}, indicating when the PT completes, i.e., eq. \eqref{Gamma} is saturated, or equivalently:
\begin{equation}\label{eq:action_max}
\frac{S_{\text{b}}}{4} +\ln{\frac{T_c}{T_n}} \approx\ln{\frac{M_{\text{Pl}}}{T_c}},
\end{equation}
where $T_n$ is the nucleation temperature for the PT to complete.

Fig. \ref{fig:nt} can be understood qualitatively. The $\epsilon$ enhancement of $S_{\text{b}}$ for prompt PT (thin wall) is absent in the asymptotic supercooled regime. Therefore, for a given $\epsilon < 1$ we have larger allowed values of $N$ for the supercooled PT than for the prompt PT. For a choice of $N$ such that the PT can complete for several values of $\epsilon$, we see that $T_n$ rises with $\epsilon$. We can understand this as follows. Supercooling can lower $S_{\text{b}}$ until the PT completes, most dramatically around $\ln(T_c/T_n)\sim 1/\epsilon$, below which the second term within the parenthesis in eq. \eqref{radionpot1} can be ignored. Furthermore, for larger $\epsilon$, $S_{\text{b}}$ is already smaller at $T\approx T_c$ and therefore less cooling is needed to complete the PT. We also see based on eq. \eqref{scaledaction} that even though the bounce action becomes $\epsilon$-independent for small temperature i.e. $\ln (T_c/T_n)\sim 1/\epsilon$, larger $\epsilon$ lets us complete the PT at higher $N$ as can be seen from eq. \eqref{eq:action_max}.

Supercooling can have important consequences. The inflationary dilution in the supercooling process $\sim (T/T_c)^3$ has to be taken into account in any baryon or dark matter genesis mechanism operating above the PT, in order to match the observed abundance today. For small $\epsilon$ we see that this is a significant issue, but not for $\epsilon \lesssim 1$ in the two-FP theory. If supercooled nucleation temperatures are very low, $T\lesssim $ GeV, in a composite Higgs context, then references \cite{vonHarling:2017yew,Baratella:2018pxi} has pointed out that QCD effects are important, for similar studies in other models, see references \cite{Witten:1980ez,Iso:2017uuu,Hambye:2018qjv}. However, in our two-FP theory with larger $\epsilon$, we see that one can have only modest cooling. Thus we neglect QCD effects. 

\section{Gravitational waves} \label{sec.gw}
Cosmological first order PTs are powerful sources of stochastic GWs. 
PTs connected very roughly to the TeV scale would produce GW amplitudes and frequencies in the range of proposed detectors. 
The GWs are produced by bubble wall collisions as well as sound waves and turbulence in the plasma (for a review see reference \cite{Caprini:2015zlo}).
While the plasma-related effects are typically expected to dominate, they are more model-dependent and less well understood than bubble collisions. However with sufficient supercooling, as we expect with small $\epsilon$, the deconfined plasma will be significantly inflated away. Below, we just consider the bubble collisions.

For bubble collisions, the peak fractional abundance and peak frequency of GWs  depend strongly on the duration of the PT, $1/\beta_{\text{GW}}$ \cite{Caprini:2015zlo},
\begin{eqnarray}
\Omega_{\text{GW}}h^2\approx 1.3\times 10^{-6}\left(\frac{H_{\text{PT}}}{\beta_{\text{GW}}}\right)^2 \left(\frac{100}{g_*}\right)^{1/3} \nonumber\\ 
f_{\text{GW}}\approx 0.04 \text{ mHz}\left(\frac{ \beta_{\text{GW}}}{H_{\text{PT}}}\right)\frac{T_{\text{PT}} }{\text{TeV}}\left(\frac{g_*}{100}\right)^{1/6},
\end{eqnarray}
where $H_{\text{PT}}$ is the Hubble scale during the PT and $\beta_{\text{GW}}$ is given by
\begin{equation}\label{beta}
\frac{\beta_{\text{GW}}}{H_{\text{PT}}}=-\frac{T}{\Gamma}\frac{d\Gamma}{dT}\bigg\rvert_{T_n} \approx -4 +T \frac{dS_{\text{b}}}{dT}\bigg\rvert_{T_n}.
\end{equation}
$h$ is defined by the present day Hubble expansion rate $H_0=100 h \text{ km/s}\text{ Mpc}^{-1}$and $g_*$ denotes the number of relativistic d.o.f. in the plasma during the PT.
$\beta_{\text{GW}}$ is (proportional to) the slope of the curves in Fig. \ref{fig:nt}, as can straightforwardly be deduced from eq. \eqref{eq:action_max} and the fact that $S_\text{b}\propto N^2$, assuming $\ln(M_{\text{Pl}}/T_c)\gg \ln(T_c/T_n)$,
\begin{equation}
\frac{\beta_{\text{GW}}}{H_{\text{PT}}} \approx -8\ln \frac{M_{\text{Pl}}}{T_c}\left(\frac{d\ln T_n}{d\ln N}\right)^{-1}.
\end{equation}

In a generic PT, $\frac{d S_{\text{b}}}{d \ln T}\sim S_{\text{b}} \sim \ln \frac{M_{\text{Pl}}}{T_c} $ . Remarkably, for small $\epsilon$, $\beta_{\text{GW}}$ is suppressed \cite{Konstandin:2011dr} and the GW abundance is enhanced. In order to see this, first note that in the supercooled regime the leading $S_{\text{b}}$ in eq. \eqref{scaledaction}, is independent of $T$. The temperature dependence arises from keeping the subleading part of the dilaton potential, eq. \eqref{radionpot1}, for small $\phi$, in the derivation of eq. \eqref{scaledaction}. This effectively results into the replacement in eq. \eqref{scaledaction} of $\lambda_0 \rightarrow \lambda_0 \left( 1-\left( \frac{T}{|\lambda_0|^{1/4}\langle\phi\rangle} \right)^\epsilon \right)$ as shown in the Appendix.  

Therefore eq. \eqref{beta} gives
\begin{equation}\label{betagw}
    \frac{\beta_{\text{GW}}}{H_{\text{PT}}} \approx -4+ 3\epsilon   \left(\frac{T_n}{|\lambda_0|^{1/4}\langle\phi\rangle}\right)^\epsilon \ln \frac{M_{\text{Pl}}}{T_c},
\end{equation}
where we have taken $\ln(M_{\text{Pl}}/T_c)\gg \ln(T_c/T_n)$.
This suppression of $\beta_{\text{GW}}$ can allow large enough GW backgrounds so that even the primordial fluctuations contained in it may be observable \cite{Geller:2018mwu}.
As $\epsilon$ increases, the PT duration decreases and bubble collision effects become less important, while the less diluted plasma effects become more important.

\section{Conclusion}\label{sec.con}
In general, (de)confinement PTs are dramatic but non-perturbative quantum phenomena. However, in this paper we have re-examined such PTs in the context of spontaneous confinement, and shown that the bubble nucleation rate is dominated by relatively simple dilaton dynamics. 
We have also shown, beyond the minimal scenario, that different near-FP regimes can control the PT dynamics and the appearance of large hierarchies, with PTs ranging from prompt to supercooled and with distinctive phenomenological features. While the detailed dynamics of deconfinement  is qualitatively important and interesting, it plays a quantitatively subdominant role in bubble nucleation.  This dynamics will be addressed in a forthcoming paper in the AdS/CFT dual context of the RS model \cite{5dpaper}.\\

\section*{Acknowledgements}
The authors would like to thank Sebastian Bruggisser, Zackaria Chacko, Benedict von Harling, Ted Jacobson, Bithika Jain, Rashmish Mishra and Riccardo Rattazzi for helpful discussions. This research was supported in part by the NSF grants PHY-1620074 and PHY-1914731, and by the Maryland Center for Fundamental Physics (MCFP). KA was also supported by the Fermilab Distinguished Scholars Program. PD was supported in part by NSF grant PHY-1915093. 
RS acknowledges the hospitality of the Kavli Institute for Theoretical Physics, UC Santa Barbara, during the ``Origin of the Vacuum Energy and Electroweak Scales'' workshop, and the support by the NSF grant PHY-174958.

\appendix
\section{Dominance of $O(3)$ symmetric, time independent bounce}
In this section, we argue that for small $\lambda_0$, the dominant dilaton bounce is $O(3)$ symmetric and independent of $1/T$ periodic Euclidean time as claimed in the main text. Our arguments will be valid for all temperatures.

Following eqs. \eqref{dilatonpot} and \eqref{radionpot1}, the generic Euclidean action for the dilaton is given by,
\begin{eqnarray} \label{eq:rescaled_action_full}
S=\frac{N^2}{4 \pi |\lambda_0|^{\frac{3}{4}}}\times &\nonumber \\\int_{0}^{1} d\tilde{t} \int dx x^2 &\left(  (\partial_x \tilde{\phi})^2+\frac{1}{|\lambda_0|^{\frac{1}{2}}}(\partial_{\tilde{t}} \tilde{\phi})^2 - \tilde{\phi}^4 \left(1-\frac{1}{1+\epsilon/4} \left(\frac{T}{|\lambda_0|^{\frac{1}{4}} \langle \phi \rangle}\right)^\epsilon \tilde{\phi}^\epsilon \right)+16 \pi^2 C \right)
\end{eqnarray}
where  $\tilde{\phi}=|\lambda_0|^{1/4}\phi/T$, $x=|\lambda_0|^{1/4} r T$, $\tilde{t}=t T$.
For simplicity, first focus on the case of  small $T$ such that $\ln(T_c/T) \gtrsim 1/\epsilon$, where the potential term proportional to  $\tilde{\phi}^{4+\epsilon}$ can be neglected. Due to the periodicity of Euclidean time and the fact that $\tilde{\phi}$ has to change by at least an $\mathcal{O}(1)$ amount in order to interpolate between the deconfined phase ($\tilde{\phi}\approx 0$) and a release point, we get $\partial_{\tilde{t}} \tilde{\phi}\sim\Delta\tilde{\phi}/\Delta \tilde{t} \sim \mathcal{O}(1)$ for a bounce profile $\tilde{\phi}$ that depends on time at the leading order. In this case, for small $\lambda_0$, due to the $\frac{1}{|\lambda_0|^{1/2}}(\partial_{\tilde{t}} \tilde{\phi})^2 $ term, a time-dependent action is parametrically larger than the time-independent $O(3)$-symmetric bounce eq. \eqref{scaledaction}. So the only way to have a smaller time-dependent bounce action, is to have a bounce that has a leading time-independent part, $\tilde{\phi}_0 (x)$, and a subleading time-dependent part, $f(x,t)$:
\begin{equation}
\tilde{\phi}(x,t)=\tilde{\phi}_0 (x)+f(x,t) ,
\end{equation}
where $f$ is of order $|\lambda_0|^{1/4}$ or smaller. The ambiguity of separating $\tilde{\phi}$ into a time dependent and time independent part is removed by requiring that $\int_{0}^{1} d \tilde{t} f=0$. In this case the action can be expanded in powers of $\lambda_0$, which to first nontrivial order in $f$ becomes
\begin{equation}
S\approx\frac{N^2}{4 \pi |\lambda_0|^{3/4}} \int_{0}^{1} d\tilde{t} \int dx x^2  \Big(  (\partial_x \tilde{\phi}_0)^2 - \tilde{\phi_0}^4+16\pi^2 C+\frac{1}{|\lambda_0|^{1/2}} (\partial_{\tilde{t}} f)^2+(\partial_x f)^2\Big),
\end{equation}
where terms linear in $f$ are not present since they vanish after integrating over $t$. The quadratic term in $f$ arising from the potential has been dropped since it has a necessarily subdominant contribution to the action for small $\lambda_0$. We see that to this order, $\tilde{\phi}_0$ and $f$ have to independently satisfy the equations of motion and a nonzero $f$ has a positive contribution to the action, so that a time-independent bounce solution has a lower action than any such time-dependent configurations/solutions. 

In the thin-wall regime, we can parallel the above arguments. In this regime the dominant contribution to the bounce comes from the region where $\phi\sim \langle\phi\rangle\gg T_c$. Thus the effective Lagrangian relevant for a thin-wall bounce can be obtained by expanding eq. \eqref{dilatonpot} around $\langle\phi\rangle$ to get,
\begin{equation}
    \mathcal{L}_{\text{eff}}=\frac{N^2}{16\pi^2}\left((\partial\phi_s)^2+2\epsilon\lambda_0\langle\phi\rangle^2\phi_s^2\right)+\cdots,
\end{equation}
where $\phi_s=\phi-\langle\phi\rangle$. Keeping the terms proportional to $\phi_s^3,\phi_s^4$ in the above expansion, will not change the parametric argument that follows.
We can recast the above using the rescalings,  $\tilde{\phi}_s=(\epsilon|\lambda_0|)^{1/4}\phi_s/T$, $\langle\tilde{\phi}_s\rangle = (\epsilon|\lambda_0|)^{1/4}\langle\phi\rangle/T_c$, $\hat{x}=(\epsilon |\lambda_0|)^{1/4} r T$, $\tilde{t}=t T$, as
\begin{equation}
   \mathcal{L}_{\text{eff}}= \frac{N^2T_c^4}{16\pi^2}\left(\frac{1}{(\epsilon|\lambda_0|^{1/2})}(\partial_{\tilde{t}}\tilde{\phi}_s)^2+(\partial_{\hat{x}}\tilde{\phi}_s)^2+2\langle\tilde{\phi}_s\rangle^2\tilde{\phi}_s^2\right)+\cdots,
\end{equation}
where we have used $T\approx T_c$ which is appropriate for the thin-wall regime. Using the above effective Lagrangian, the Euclidean action can be constructed. Then we can repeat all the arguments given above for the supercooling regime to conclude again that the dominant bounce is time-independent, this time due to the smallness of the quantity $\epsilon\lambda_0$. Even for intermediate $T$, these arguments can be generalized to show time-independence of the dominant dilaton bounce.

%
%
\section{Subleading temperature correction to the bounce action in the supercooled regime } 
In this section we calculate the subleading correction to $S_{\text{b}}$ in the supercooled  regime, using which we can find the parameter $\beta_{\text{GW}}$ relevant for gravitational waves as in eq. \eqref{betagw}. The relevant action can be read off from eq. \eqref{eq:rescaled_action_full} by dropping the time-dependent contribution,
\begin{equation} \label{eq:rescaled_action}
S_{\text{b}}=\frac{N^2}{4 \pi |\lambda_0|^{3/4}} \int dx x^2 \left(  (\partial_x \tilde{\phi})^2 - \tilde{\phi}^4 \left(1-\frac{1}{1+\epsilon/4} \left(\frac{T}{|\lambda_0|^{1/4} \langle \phi \rangle}\right)^\epsilon \tilde{\phi}^\epsilon \right)+16 \pi^2 C \right).
\end{equation}
We will expand in $ \left(\frac{T}{|\lambda_0|^{1/4} \langle \phi \rangle}\right)^\epsilon$ by treating the term in the potential proportional to $\tilde{\phi}^{4+\epsilon}$ as a perturbation, and obtain the leading temperature correction to $S_{\text{b}}$ by first solving the ``zeroth-order'' bounce equation in the absence of the $\tilde{\phi}^{4+\epsilon}$ term. Let us denote such a bounce solution as $\tilde{\phi}_0(x)$ and the corresponding zeroth-order bounce action as $S_{\text{b}}^{(0)}$. The leading correction to $S_{\text{b}}^{(0)}$ is then given by
\begin{equation}
\Delta S_{\text{b}}=\frac{N^2}{4 \pi |\lambda_0|^{3/4}} \frac{1}{1+\epsilon/4} \left(\frac{T}{|\lambda_0|^{1/4} \langle \phi \rangle}\right)^\epsilon \int dx x^2 \tilde{\phi}_0^{4+\epsilon}.
\end{equation}
Note that even though the solution $\tilde{\phi}$ is corrected by the perturbation, the change of the action due to this correction vanishes to first order since the first variation of the action vanishes when evaluated on the solution of equation of motion.
Then, for small $\epsilon$, we can approximate the above correction as,  
\begin{equation}
\Delta S_b\approx \frac{N^2}{4 \pi |\lambda_0|^{3/4}} \left(\frac{T}{|\lambda_0|^{1/4} \langle \phi \rangle}\right)^\epsilon\int dx x^2   \tilde{\phi}_0^{4}.
\end{equation}
This implies the temperature dependent bounce action can be approximated as,
\begin{eqnarray} \label{eq:rescaled_action}
S_{\text{b}}\approx&\frac{N^2}{4 \pi |\lambda_0|^{3/4}} \int dx x^2 \left(  (\partial_x \tilde{\phi}_0)^2 - \tilde{\phi}_0^4 \left(1-\frac{1}{1+\epsilon/4} \left(\frac{T}{|\lambda_0|^{1/4} \langle \phi \rangle}\right)^\epsilon  \right)+16 \pi^2 C \right)\\
\approx&\frac{N^2}{4\pi T}\int dr r^2 \left( \left(\frac{d\phi}{dr}\right)^2+\lambda_0\left(1-\left(\frac{T}{|\lambda_0|^{1/4} \langle \phi \rangle}\right)^\epsilon\right)\phi^4 +16 \pi^2 CT^4\right),
\end{eqnarray}
where in the second line we have re-expressed the action in terms of $\phi$ and $r$.
Therefore including this subleading temperature correction is equivalent to a corresponding change in the dilaton quartic coupling,
\begin{equation}
\lambda_0\rightarrow\lambda_0\left(1-\left(\frac{T}{|\lambda_0|^{1/4} \langle \phi \rangle}\right)^\epsilon\right)
\end{equation}
in eq. \eqref{scaledactionfull}.
Thus to first order in $\left(\frac{T}{|\lambda_0|^{1/4} \langle \phi \rangle}\right)^\epsilon$ and for small $\epsilon$ we have using eq. \eqref{scaledaction},
\begin{equation}
S_{\text{b}} \approx S_{\text{b}}^{(0)} \left(1+ \frac{3}{4} \left(\frac{T}{|\lambda_0|^{1/4} \langle \phi \rangle}\right)^\epsilon  \right),
\end{equation}
which when used in eq. \eqref{beta} gives eq. \eqref{betagw}.

\bibliographystyle{JHEP}
\bibliography{refs_RSPT}

\end{document}